\documentclass[5p,twocolumn,number]{elsarticle}

\usepackage{epsfig}

\begin{document}

\begin{frontmatter}

\title{A measurement method of a detector response function for monochromatic electrons based on the Compton scattering.}

\author[pnpi]{S.V.~Bakhlanov}

\author[spsu]{N.V.~Bazlov}

\author[pnpi]{A.V.~Derbin}
 \ead{derbin@pnpi.spb.ru}

\author[pnpi,gssi]{I.S. Drachnev}

\author[pnpi]{A.S. Kayunov}

\author[pnpi]{V.N.~Muratova}

\author[pnpi]{D.A.~Semenov}

\author[pnpi]{E.V.~Unzhakov}

\address[pnpi]{St.Petersburg Nuclear Physics Institute,National Research Center "Kurchatov Institute",
 Gatchina, Russia 188300}
%\address[spsu]{ Saint-Petersburg State University, Universitetskaja nab. 7/9, Saint-Petersburg, Russia, 199034}
\address[spsu]{V.A. Fok Institute of Physics, St. Petersburg State University, Russia, 199034}
\address[gssi]{GranSasso Science Institute, INFN, L'Aquila (AQ) I-67100, Italy}

\begin{abstract}
In this paper we present a method of scintillation detector energy calibration using the gamma-rays.  The technique is
based  on the Compton scattering of gamma-rays in a scintillation detector and subsequent photoelectric absorption of
the scattered photon in the Ge-detector. The novelty of this method is that the source of gamma rays, the germanium and
scintillation  detectors are immediately arranged  adjacent to each other. The method presents an effective solution
for the detectors consisting of a low atomic number materials, when the ratio between Compton effect and photoelectric
absorption is large and the mean path of gamma-rays is comparable to the size of the detector. The technique can be
used for the precision measurements of the scintillator light yield dependence on the electron energy.
\end{abstract}

\begin{keyword}
 scintillator and semiconductor detectors \sep energy scale calibration

% \PACS 14.80.Mz \sep 29.40.Mc \sep 26.65.+t
\end{keyword}

\end{frontmatter}

\section{Introduction}
%%%%%%%%%%%%%%%%%%%%%%%%%%%%%%%%%%%%%%%%%%%%%%%%%%%%%%%%%%%%%%%%%%%%%%%%%%%%%%%%%%%%%%%%%%%%%%%
A wide range of detectors is difficult to calibrate using gamma or X-rays, because of the absence of full absorption peaks in the
spectrum. In particular, it includes the detectors with spatial dimensions that are relatively small compared to the mean path of
calibration source gamma rays, and also the ones which consist of a material with a low atomic number. As a result, the
probability of the photoelectric effect is small compared to the probability of the Compton scattering, and a small detector
dimensions make it impossible to obtain a full absorption peak which could appear due to the multiple scattering. The calibration
by using the edges of the Compton scattering has a significantly lower accuracy.

Common examples include  widespread plastic and liquid hydrocarbon scintillators $(\rm{C_n H_m})$ or  semiconductor detectors of
a large volume based on silicon. For example, the Compton scattering cross-section of gamma rays at 0.1 (1.0) MeV on carbon is
$\sigma_{C}$ = 2.9 (1.3) barn/atom, and the photoelectric effect cross-section is lower by two orders of magnitude --
$\sigma_{ph}$ = 0.021 $(2.7\times10^{-5})$ barns/atom. The corresponding  gamma ray Compton scattering mean paths are
${0.15~\rm{g/cm^{2}}}$ (7 cm at a density of ${1~\rm{g/cm^3}}$) and $0.064~\rm{g/cm^2}$ (16 cm). As a result, the scintillation
volume is mostly populated by the Compton-produced electrons, and therefore it would be preferable to use those electrons for
calibration.

A classical approach \cite{Hof50, Sie79} utilizes a narrowly collimated beam of gamma-rays and mutually spaced
scintillation  detectors wired into the coincidence unit. This design allows one to  identify clearly the scattering
angle and determine the amount of energy deposited in the scintillator. The disadvantages of this method are that it
requires an intense source of gamma-rays, and the data acquisition is quite time-consuming requiring a separate
measurement for each particular value of  scattered electron energy.

The key feature of our method is that the source of gamma-rays, and both scintillation and semiconductor detectors are located
close to each other and this fact makes it possible to  record simultaneously coincidence events in a large range of scattering
angles , and thus to measure the scintillator light yield for the Compton's electrons within a wide energy range.

\section{Experimental setup}
\begin{figure}
\includegraphics[width=8cm,height=7.5cm, bb = 20 100 500 700]{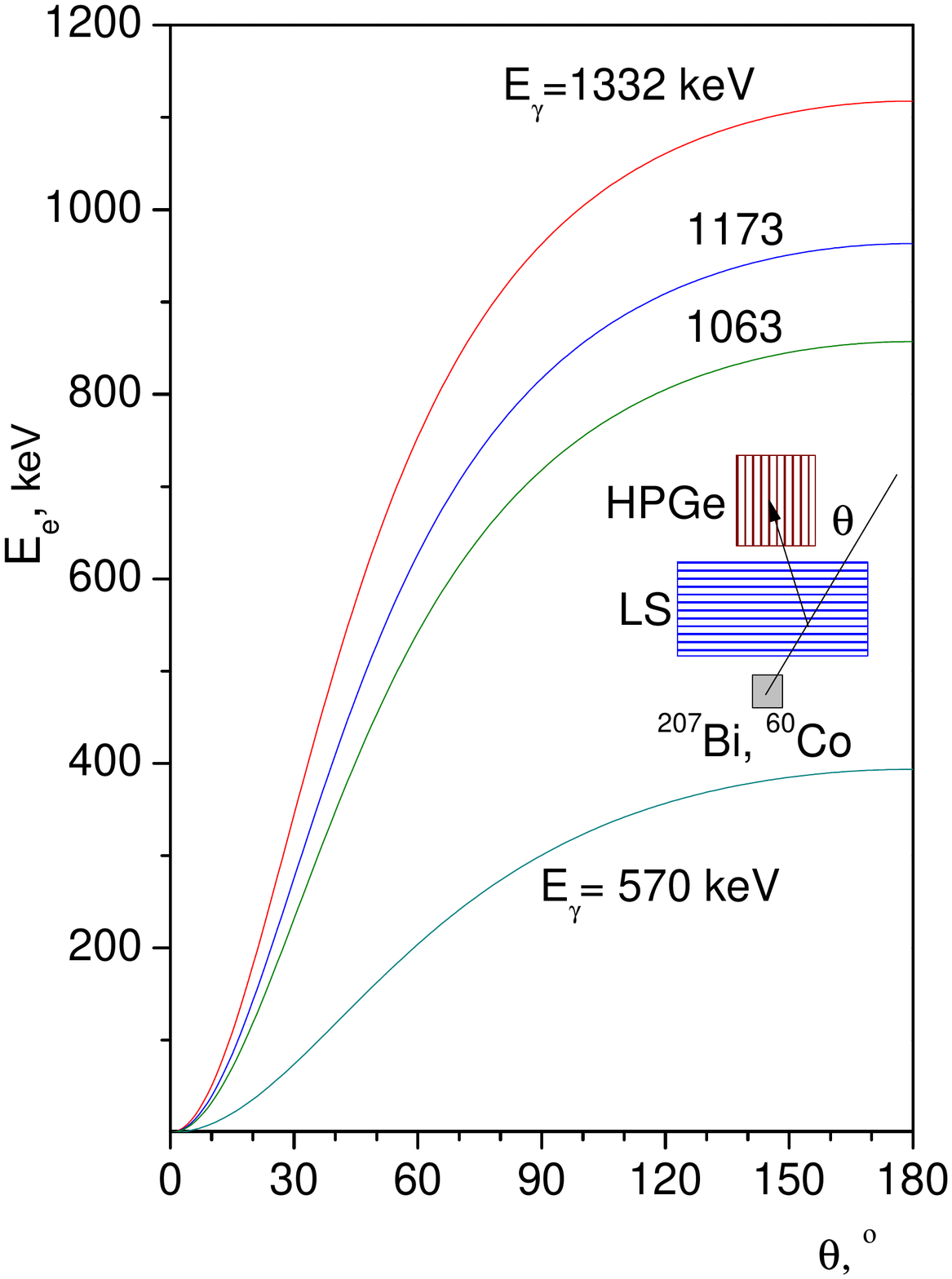}
\caption {The dependence of the electron Compton energy ${E}_e$ on the scattering angle for  different gamma-ray
sources ($^{60}$Co and $^{207}$Bi). The inset shows the layout of the experimental setup.} \label{fig1}
\end{figure}

%%%%%%%%%%%%%%%%%%%%%%%%%%%%%%%%%%%%%%%%%%%%%%%%%%%%%%%%%%%%%%%%%%%%%%%%%%%%%%%%%%%%%%%%%%%%%%%%%%%%%%%%%%%%%%%%%%%%%%%
The experimental setup with a liquid scintillation (LS) and semiconductor (HPGe) detectors is shown on the inset of
Fig. \ref{fig1}. This layout allows one to tell the coincidence events of Compton scattering in the scintillator from
the photoelectric absorption events in Ge-detector.

The sources with an activity of about $10^5$ Bq (included in the standard reference source kits) are sufficient to
obtain the required statistics, while the measurements are carried out at the same time for a range of electron
energies that is limited by the maximum scattering angle. We used two reference sources made of $^{60}$Co (energies
1173 keV and 1333 keV) and $^{207}$Bi (gamma lines at 570 keV and 1063 keV). For example, provided that the
$^{60}$Co-source, LS- and Ge-detectors are all aligned with the maximum scattering angle of $90^{\circ}$, one should be
able to study the recoil electrons energy within the 0 - 950 keV range (Fig. \ref{fig1}).

In order to register the scattered gamma-rays we used a cylindrical HPGe-detector with the diameter of 60 mm and length
of 60 mm, which had an energy resolution FWHM (full width at half maximum) of 1.0 keV at 122 keV (gamma line of a
$\rm{^{57}Co}$ source). The detector was connected to the standard spectrometer channel, consisting of a
charge-sensitive preamplifier with the resistive feedback and cooled Field Effect Transistor (FET), spectrometric
amplifier with $2~\mu s$ shaping time and 12-bit analog-to-digital converter (ADC). The gain value was chosen for the
ADC energy scale to be equal to 0.4 keV/channel.

\begin{figure}
\includegraphics[width=8cm,height=6cm, bb = 0 0 180 130]{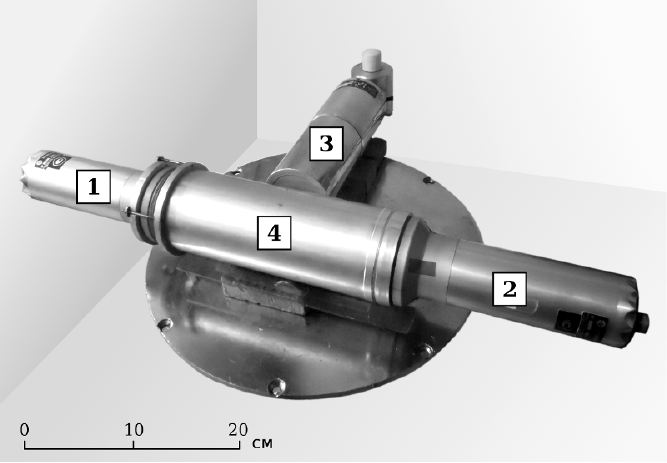}
\caption {The photo of the experimental setup: 1,2 - two PMTs, 3 - HPGe-detector, 4 - LS-detector} \label{fig2}
\end{figure}

The LS-detector is designed as a hollow teflon (PTFE) cylinder with the internal diameter of 60 mm and  length of 100
mm, which is filled with a liquid scintillator. Two photomultiplier tubes (PMTs) R1307 are attached to the both ends of
the cylinder.  The PMTs are separated from the scintillator volume by 10-mm disc plates made of acrylic, which prevents
the generation of events near the surface of photocathode. Linear alkylbenzene (LAB) supplemented with PPO
concentration 0.5 g/l \cite{Nem11} was used as a scintillator. Each PMT had a dedicated spectrometer channel consisting
of a preamplifier, an amplifier and a 10-bit ADC. The amplitude ratio between  signals from  PMT1 and PMT2 allows one
to select a fiducial volume in the central region of the LS-detector. The photo of HPGe- and LS-detector is shown in
Fig. \ref{fig2}.

The coincidence circuit was implemented using the ${\rm{CAMAC}}$ standard. As a result, the computer memory accumulated
the sequentially recorded amplitude values for three signals, corresponding to HPGe detector coincidences with PMT1 and
PMT2 of the LS-detector. The stored data file was then processed off-line, once the sufficient statistics was obtained.
A three dimensional spectrum is formed during the data collection: the number of events N, the amplitude of the
Ge-detector and the sum of the PMT1 and PMT2 amplitudes. For monitoring purposes a three-dimensional plot is formed
on-line during the data acquisition, picturing the number of events N versus the amplitude of Ge-detector and sum of
the amplitudes from PMT1 and PMT2.

\section{Results}
\begin{figure}
\includegraphics[width=8cm,height=7.5cm, bb = 20 100 500 700]{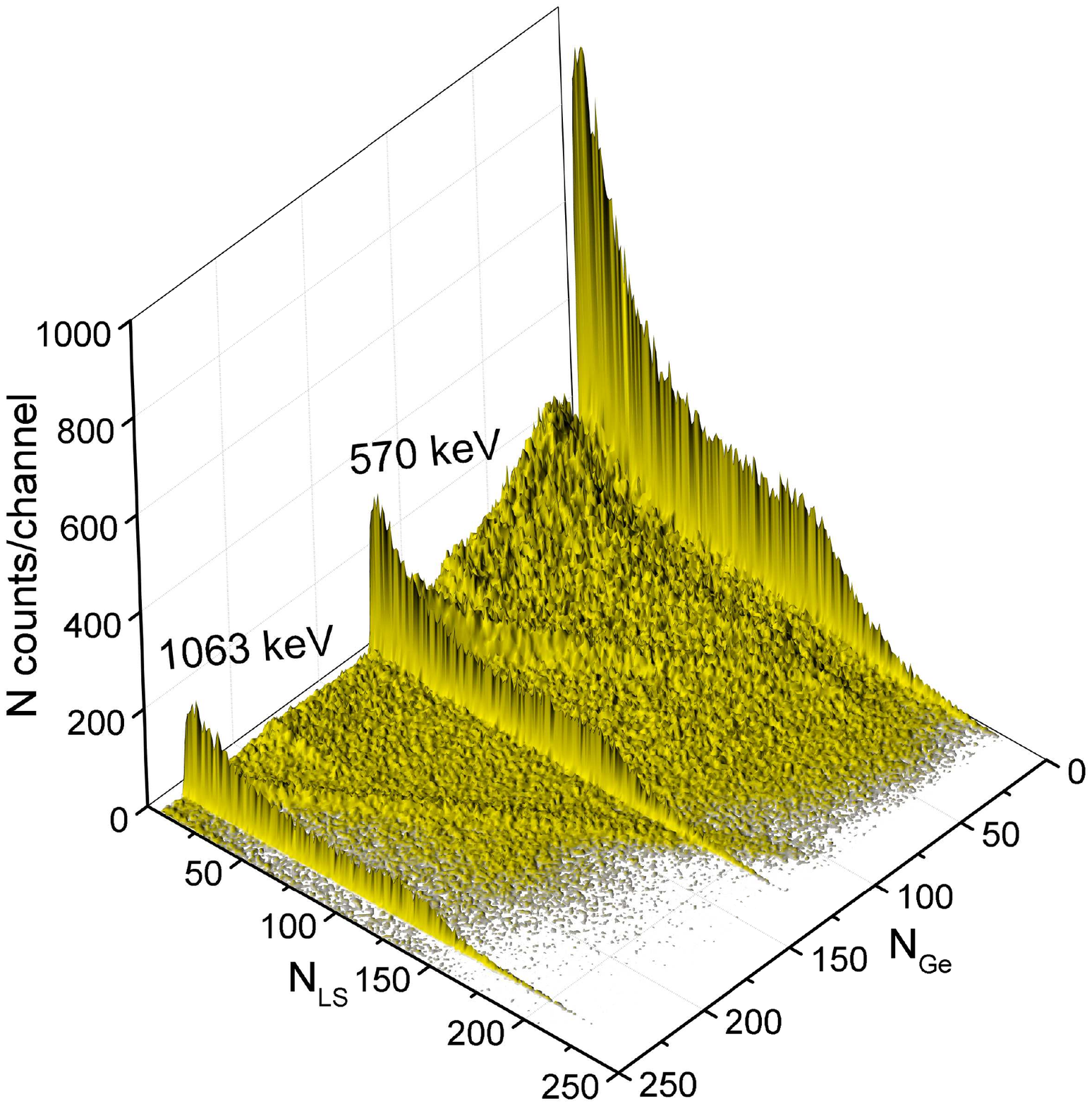}
\caption {The 3D plot for the $^{207}Bi$ source measurement.} \label{fig3}
\end{figure}

%%%%%%%%%%%%%%%%%%%%%%%%%%%%%%%%%%%%%%%%%%%%%%%%%%%%%%%%%%%%%%%%%%%%%%%%%%%%%%%%%%%%%%%%%%%%%%%%%%%%%%%%%%%%%%%%%%%%%%%%%%
A 3D plot for a series of measurements with the $^{207}$Bi source is shown in Fig. \ref{fig3}. The ${N}_{LS}$ and
${N}_{Ge}$ axes correspond  to the amplitudes of the LS-detector and HPGe-detector accordingly. The figure clearly
shows two linear ridges, crossing the (${N}_{Ge},{N}_{LS}$) plane, which start at the energies of ${E_\gamma = 570}$
and ${E_\gamma = 1063}$ keV and are governed by relation ${E_\gamma} = {E}_{Ge} + {E}_{LS}$.  Those ridges correspond
to the Compton scattering in the scintillator and the subsequent total absorption of the scattered gamma quantum in the
Ge-detector. These features therefore allow us to  select the Compton electrons of the given energy.

\begin{figure}
\includegraphics[width=8cm,height=8.5cm, bb = 20 100 500 700]{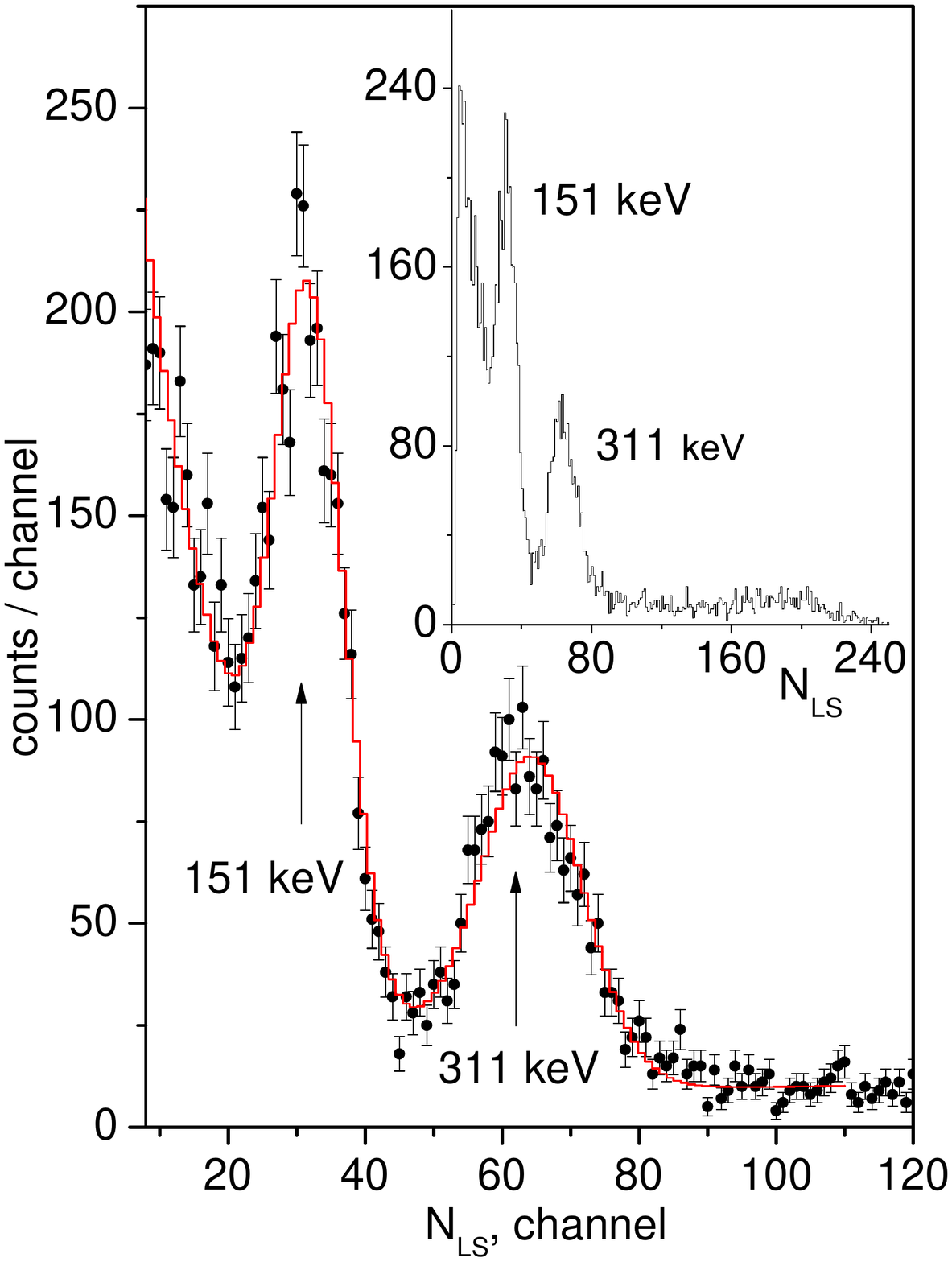}
\caption {LS-detector spectrum from the $^{60}$Co source for the Ge-detector energy deposit value of $1022\pm 2$ keV. The inset
shows the spectrum in wide energy range.} \label{fig4}
\end{figure}

For each energy value ${E}_{Ge}^i$ which was deposited within the Ge-detector and stored in the channel i, there is a
peak in the spectrum of LS-detector, that corresponds to the energy of the Compton electron ${E}_{LS}^i = {E}_\gamma -
{E}_{Ge}^i$. Fig. \ref{fig4} shows the spectrum of LS-detector, measured with a $^{60}$Co source, corresponding to the
energy deposit of 1022 keV inside the Ge-detector. There are two peaks formed by the scattering of $^{60}$Co lines at
1173 keV and 1332 keV. Accordingly, the peaks in the LS-spectrum are located at 151 keV and 311 keV.

The position of the LS-detector peak ${N}_{LS}^i$ is determined by fitting  the spectrum using the sum of the Gaussian
and the baseline function selected to describe the continuous background (Fig. \ref{fig4}). In fact, it is sufficient
to process a few peaks in order to determine the energy scale of LS-detector and the energy dependence of the
resolution.

\begin{figure}
\includegraphics[width=8cm,height=8.5cm, bb = 20 100 500 700]{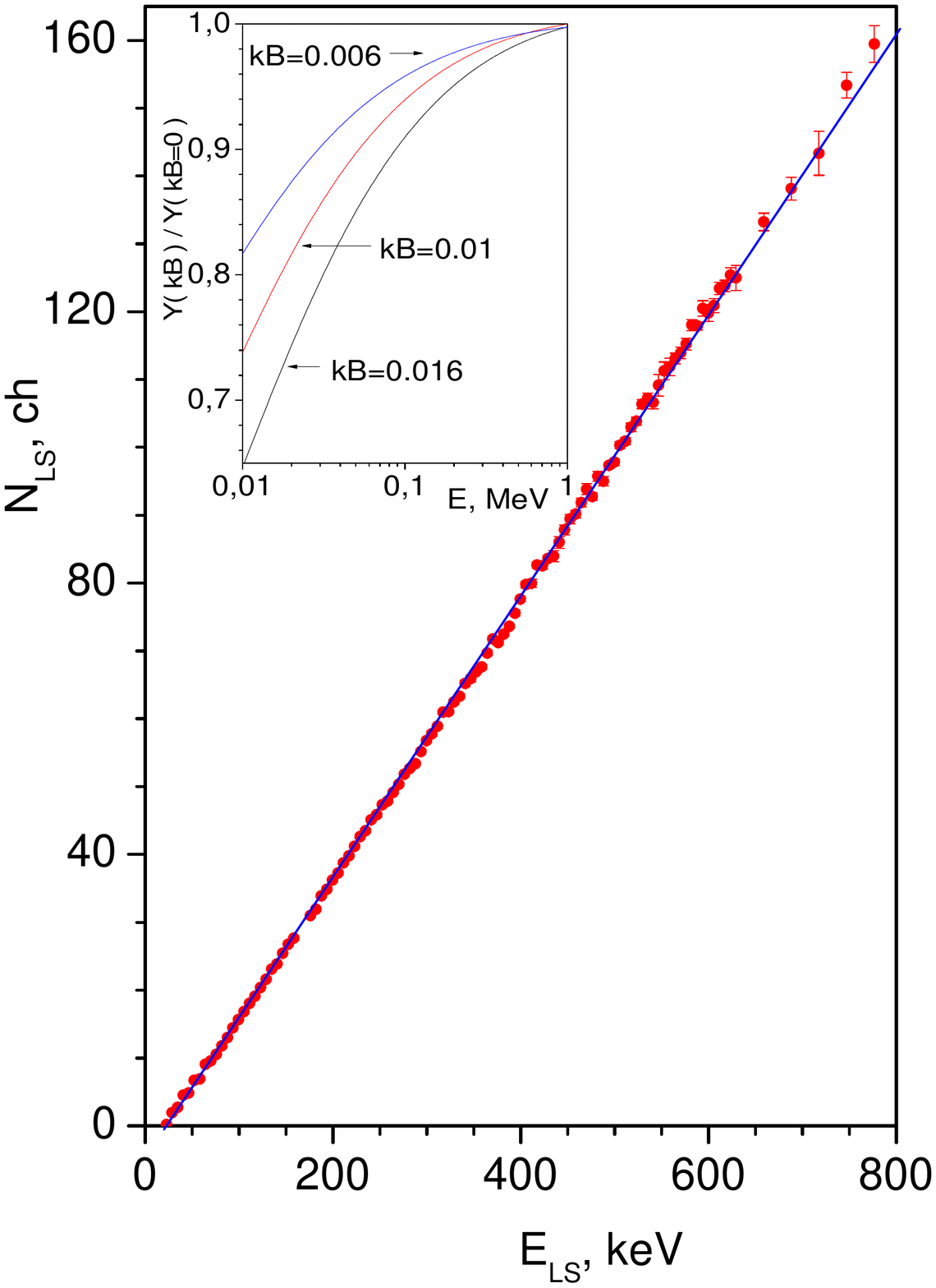}
\caption {The light yield dependence on the energy of Compton electron. The inset shows the Y(kB)/Y(kB=0) dependence on the
electron energy for different kB values.} \label{fig5}
\end{figure}

Modern computational capabilities make it possible to process all of the obtained spectra. The amplitude detector
LS-electron energy dependence in a certain range (20 - 800) keV for 120 points, is shown in Fig. \ref{fig5}. It should
be noted that when using a point source at the fixed position, this method requires the uniformity of light collection
as different scattering angles correspond to different regions of the scintillator. However, this circumstance is not
essential due to the poor energy resolution of scintillation detectors.

Since the amplitude of the scintillation detector${N}_{LS}$ depends on the light output (Y), which is a fundamental
property  of the scintillator, this method allows us to study the  dependence of the light yield on the electron energy
Y(E). In particular, the deviation  of Y(E) from a linear dependence at small electron energies ($\leq300$ keV),is
called ionization quenching effect. A precise measurement of Y(E) is important for a number of physical problems in
which the obtained scintillation spectrum is  fitted by a theoretical electron energy spectrum.

The deviation from the linear dependence between the number of photons dY per length unit dx and energy losses dE/dx is
usually described by the phenomenological Birks relation \cite{Bir51, Bir75}:

\begin{equation}
\frac{dY}{dx} = {{Y}_0}\frac{dE/dx}{1+kB\cdot{dE/dx}}.
\end{equation}

Where kB is the Birks coefficient (a quenching factor) and ${Y}_0$ is the light yield for kB = 0. These two values are inherent
characteristics of the particular scintillator. The typical values are kB = 0.01 cm/MeV and ${Y}_0$ = $10^4$ photons/MeV
\cite{Bel14}. The Y(E,kB) dependence is obtained by integrating the expression (1) over the energy, taking into account the
energy losses dE/dx for a particular substance. In case of LAB ($\rm{C_{18}H_{30}}$) the obtained Y(kB)/Y(kB=0) dependence on the
electron energy for different values of kB are shown in the inset of Fig. \ref{fig5}. The ratios are normalized to unity at E = 1
MeV. It can be seen that at the energy of 100 keV the deviation of Y(kB)/Y (kB = 0) from 1 is (4-9)\% for kB = (0.006, 0.01,
0.016).

The method described above allows us to measure the entire Y(E) dependence, which can be used instead of the relation
(1), which is determined only by two parameters kB and ${Y}_0$. In our case, to achieve the position accuracy of less
than 1\% for the 100 keV peak, the statistics has to be greater than ${10}^3$ events in the peak (Fig. \ref{fig4}).

There are two basic problems in this approach (which are also present in the classical approach) which are associated
with the multiple Compton scattering (mostly double scattering) in the LS-detector and the non-uniformity of light
collection within the liquid scintillator volume. The mean path of gamma-ray at 1.33 MeV ($^{60}$Co) is 20 cm for LAB.
For the described LS-detector the probability of double scattering at small angles, which represent the main interest
in measurements of Y(E) dependence, is 15\%. The impact of both effects can be accounted for by performing the
Monte-Carlo simulations.

\newpage
\section{Conclusion}
%%%%%%%%%%%%%%%%%%%%%%%%%%%%%%%%%%%%%%%%%%%%%%%%%%%%%%%%%%%%%%%%%%%%%%%%%%%%%%%%%%%%%%%%%%%%%%%
The method of calibration of the scintillation detector energy scale with a gamma-ray based on the Compton scattering
in the  scintillation detector and  subsequent total absorption of the scattered photon in Ge-detector is proposed. The
gamma-ray source, scintillation detector and  Ge-detector are located almost next to each other. The calibration method
is effective for the detectors made of  materials with a low atomic number or when the dimensions of the detector are
comparable with the length of a gamma-ray mean path. The technique can be used to study the dependence of the
scintillator light yield on the energy of electrons.

This work was supported by the Russian Foundation for Basic Research (projects  15-02-02117-A, 13-02-01199-A and
13-02-12140-ofi-m).

%%%%%%%%%%%%%%%%%%%%%%%%%%%%%%%%%%%%%%%%%%%%%%%%%%%%%%%%%%%%%%%%%%%%%%%%%%%%%%%%%%%%%%%%%%%%%%%

\end{document}